\documentclass[doublespacing]{elsart}
\usepackage{graphicx}

\input{tcilatex}
\begin{document}

\begin{frontmatter}
\title{Tsallis distribution  and luminescence decays}
\author{Kwok Sau Fa}
\ead{kwok@dfi.uem.br}

\address{Departamento de F\'{\i}sica, Universidade Estadual de Maring\'{a}, Av.
Colombo 5790, 87020-900, \ Maring\'{a}-PR, Brazil, Tel.: 55 44 32614330,
Fax: 55 44 32634623}

\begin{abstract}
Usually, the Kohlrausch (stretched exponential) function is employed to fit the luminescence decays. 
In this work we propose to use the Tsallis distribution as an alternative to describe them.
 We show that the curves of the luminescence decay obtained from the Tsallis distribution are close to 
those ones obtained from the stretched exponential. 
Further, we show that our result can fit well the data of porous silicon at low temperature and simulation 
result of the trapping controlled luminescence model.
\end{abstract}
\end{frontmatter}

\section{Introduction}

The Kohlrausch (stretched exponential) function has been largely employed to
describe luminescence decays of different materials and in different time
scales (see \cite{chen,lee,xchen,berberan1,berberan2,berberan3,pavesi}, and
references therein), and it has the following form:

\begin{equation}
I_{\beta }(t)=I_{\beta 0}\exp \left[ -\left( \frac{t}{\tau }\right) ^{\beta }%
\right] \text{ ,}  \label{eq1}
\end{equation}%
with $0<\beta <1$. We note that the value of parameter $\beta $ in the range 
$0<\beta <1$ makes $I(t)$ decays slower than the exponential function,
whereas for $\beta >1$ the function (\ref{eq1}) is called compressed
exponential and it has short-tails, consequently, it decays faster than the
exponential function.

For the general consideration of the study of luminescence decay $I(t)$, it
is usually considered the distribution of rate constants $H(k)$ \cite%
{berberan1} connected with $I(t)$ by the Laplace transform given by%
\begin{equation}
I(t)=\int_{0}^{\infty }H(k)\text{e}^{-kt}\text{d}k\text{ ,}  \label{eq2}
\end{equation}%
with $I(0)=1$. In particular, $H(k)$ must be nonnegative for all $k>0$ in
order to be considered as a distribution function. Moreover, the function $%
H(k)$ is normalized, i.e., $\int_{0}^{\infty }H(k)$d$k=1$. We note that $%
H(k) $ may have a large change for a small change in $I(t)$. In this way,
the precision of the experimental data will be important for the choice of
the distribution of rate constants $H(k)$.

Modification and generalization of the stretched exponential function as a
decay function has been recently considered, for instance, a decay function
unifying the modified stretched exponential and Becquerel decay laws \cite%
{berberan1,berberan2,berberan3}.

In this work we consider a simple distribution of rate constants $H(k)$
based on the Tsallis distribution \cite{tsallis1,tsallis2}. Our idea of
employing this distribution is due to its simplicity and it has been applied
to a variety of natural systems (see, for instance, \cite%
{plastino1,plastino2,lyra,beck}). We shall show that the luminescence decay
function generated by the Tsallis distribution can be useful to describe
experimental data. To do so we shall fit the data of porous silicon at low
temperature \cite{xchen} and the simulation result of the trapping
controlled luminescence model described in \cite{chen} with our luminescence
decay function.

\bigskip

\section{Distribution of rate constants $H(k)$ given by the Tsallis
distribution}

For the luminescence decay $I(t)$ given by Eq. (\ref{eq1}) the distribution $%
H(k)$ can be expressed in an integral representation \cite{berberan1} as
follows:

\begin{equation}
H_{\beta }(k)=\frac{\tau _{0}}{\pi }\int_{0}^{\infty }du\exp \left[
-u^{\beta }\cos \left( \frac{\beta \pi }{2}\right) \right] \cos \left[
u^{\beta }\sin \left( \frac{\beta \pi }{2}\right) -k\tau _{0}u\right] \ .
\label{eq4}
\end{equation}%
We see that this last expression is quite complicated. Now we consider the
Tsallis distribution for $H(k)$, and it is given by

\begin{equation}
H_{q}(k)=B\left[ 1-(1-q)\alpha k\right] ^{\frac{1}{1-q}}  \label{eq5}
\end{equation}%
where $\alpha >0$ and $0<q<2$, and $B$ is a normalization factor. The
parameter $\alpha $ has the dimension of time. For $q\rightarrow 1$ we
recover from Eq. (\ref{eq5}) the exponential function which is connected
with a particular case of \ the Becquerel decay function \cite{berberan3}
given by

\begin{equation}
I(t)=\frac{1}{\left( 1+\frac{\gamma t}{\alpha }\right) ^{\frac{1}{\gamma }}}%
\text{ .}  \label{eq5a}
\end{equation}%
We note that $H_{q}(k)$ has a cutoff for $q<1$, i.e., the term $\left[
1-(1-q)\alpha k\right] ^{\frac{1}{1-q}}$ is replaced by zero when $\left[
1-(1-q)\alpha k\right] <0$, then the normalization factor is equal to 
\begin{equation}
B=\alpha (2-q)  \label{eq6}
\end{equation}%
where $0<q<2$. The luminescence decay $I(t)$ is obtained by substituting Eq.
(\ref{eq5}) into Eq. (\ref{eq2}), and we arrive at 
\begin{equation}
I_{q}(t)=\frac{2-q}{1-q}e^{-\frac{t}{\alpha (1-q)}}\int_{0}^{1}du\ u^{\frac{1%
}{1-q}}e^{\frac{tu}{\alpha (1-q)}}\text{ , \ }0<q<1\text{,}  \label{eq7}
\end{equation}%
and%
\begin{equation}
I_{q}(t)=\frac{2-q}{q-1}e^{-\frac{t}{\alpha (q-1)}}\int_{1}^{\infty }du\ u^{%
\frac{1}{1-q}}e^{\frac{tu}{\alpha (1-q)}}\text{ , \ }1<q<2\text{.}
\label{eq7a}
\end{equation}%
In order to see the behavior of $I_{q}(t)$ we plot some curves of $I_{q}(t)$
\ and $I_{\beta }(t)$ with typical values of $\alpha $, $q$ and $\beta $
which are shown in Fig. 1. All the curves decay monotonically with time, and
two of the curves of $I_{q}(t)$ have fat tails. We note that the lowest
curve of $I_{q}(t)$ in the figure is very close to the curve of $I_{\beta
}(t)$; This means that any experimental data fitted with these curves must
have a good level of precision in order to choose which one of the curves is
preferable.

For application of our result we consider the experimental data of porous
silicon at low temperature given in \cite{xchen}. In Fig. 2 we compare the
best fit of the data by using the stretched exponential and $I_{q}(t)$. We
see that both curves can fit the data (which are not shown in our figure)
very well. Then, in this case, the experimental data do not offer us
sufficient precision to discard one of the curves. We note that the behavior
of $I_{q}(t)$ is not the same of \ $I_{\beta }(t)$, i.e., $I_{\beta }(t)$
describes a straight line, whereas $I_{q}(t)$ does not.

Another application of our result is to consider the model of trapping
controlled luminescence given in Ref. \cite{chen}. The process during the
excitation is described by the following equations:

\begin{equation}
\frac{\text{d}n_{\nu }}{\text{d}t}=x-R\left( M-m\right) n_{\nu }
\label{eq8a}
\end{equation}%
\begin{equation}
\frac{\text{d}m}{\text{d}t}=-A_{m}mn_{c}+R\left( M-m\right) n_{\nu }
\label{eq8b}
\end{equation}%
\begin{equation}
\frac{\text{d}n}{\text{d}t}=A_{n}\left( N-n\right) n_{c}  \label{eq8c}
\end{equation}%
\begin{equation}
\frac{\text{d}n_{c}}{\text{d}t}=\frac{\text{d}m}{\text{d}t}+\frac{\text{d}%
n_{\nu }}{\text{d}t}-\frac{\text{d}n}{\text{d}t}\text{ ,}  \label{eq8d}
\end{equation}%
where $n_{c}$ and $n_{\nu }$ are the instantaneous concentrations of
electrons in the conduction band and holes in the valence band,
respectively; $A_{n}$ is the retrapping coefficient, $A_{m}$ is the
recombination coefficient and $R$ is the trapping coefficient of free holes
during the excitation; $N$ and $M$ are the concentrations of the traps and
recombination centers, whereas $n$ and $m$ are their respective
instantaneous occupancies. Finally $x$ is the rate of production of
electrons and holes by the excitation irradiation. The luminescence emission
intensity is calculated by the rate of electron-hole recombination given by%
\begin{equation}
I=-\frac{\text{d}m}{\text{d}t}=-A_{m}mn_{c}\text{ .}  \label{eq9}
\end{equation}

In Fig. 3 shows a replotting of the simulation result \cite{chen} and the
best fitted stretched exponential function with the parameters $x=10^{19}$m$%
^{-3}$s$^{-1}$, $A_{m}=10^{-17}$m$^{3}$s$^{-1}$, $A_{n}=10^{-9}$m$^{3}$s$%
^{-1}$, $R=10^{-17}$m$^{3}$s$^{-1}$, $N=10^{18}$m$^{-3}$, $M=10^{19}$m$^{-3}$%
. We note that the agreement is not very good. In order to enhance the
fitting result the decay curve has been separated into two parts \cite{chen}%
: The values have been fitted separately for the first microsecond and for
the period of time from $t=2$ to $10\mu $s. \ In Fig. 4 shows the simulation
result fitted by our result; we see that the agreement is excellent and the
decay curve has not been separated.

\bigskip

\section{Conclusion}

\bigskip

In this work we have considered the Tsallis distribution as a distribution
of rate constants $H(k)$. From Eq. (\ref{eq2}) we have obtained the
luminescence decay function $I_{q}(t)$ for fitting the luminescence decays.
We have shown that $I_{q}(t)$ may have fat tails and its behavior looks like
the stretched exponential function. We have also demonstrated that $I_{q}(t)$
can be useful to describe experimental data. In the case of the trapping
controlled luminescence model the fitted result by $I_{q}(t)$ is superior
than the best fitted stretched exponential function. It can be seen that,
Fig. 3, the deviation of the best fitted stretched exponential function from
the exponential one is not small $\beta =0.45$. On the other hand, Fig. 4
shows the simulation result fitted by $I_{q}(t)$ with $q=1.04$ which
presents a small deviation of $H_{q}(k)$ from the exponential function, then 
$H_{q}(k)$ is close to the Becquerel distribution of rate constants $\alpha
\exp \left( -k\alpha \right) $; however, the parameter $\alpha =13.0321\mu $%
s has \ a value close to $\tau =11.8\mu $s.

\textbf{Acknowledgment}

The author acknowledges partial financial support from the Conselho Nacional
de Desenvolvimento Cient\'{\i}fico e Tecnol\'{o}gico (CNPq), Brazilian
agency.

\newpage

\begin{center}
\textbf{Figure Captions}

\bigskip

\bigskip

\bigskip
\end{center}

Fig. 1 - \ Plots of $I_{q}(t)$ and $I_{\beta }(t)$ given by Eqs. (\ref{eq1})
and (\ref{eq7a}) in arbitrary units. The dotted lines correspond to $%
I_{q}(t) $, whereas the solid lines correspond to $I_{\beta }(t)$ with
typical parameter values: from top to bottom, $\beta =0.6$, $\tau =4.56$, $%
\alpha =(1.64)^{2}$ , $q=1+(0.28)^{2}$ ; $\beta =0.65$, $\tau =3.07$, $%
\alpha =(1.4)^{2}$ , $q=1+(0.35)^{2}$ ; $\beta =0.45$, $\tau =0.68$, $\alpha
=(0.695)^{2}$ , $q=1+(0.275)^{2}$.

Fig. 2 - The best fit to Eq. (\ref{eq1}) (solid line) for porous silicon at
low temperature described in \cite{xchen} with $\beta =0.75$ and $\tau =2.4$%
ms. The dotted line is the plot of Eq. (\ref{eq7a}) with $q=1+(0.25)^{2}$
and $\alpha =(1.235)^{2}$ms.

\bigskip

Fig. 3 - Replotting of the simulation result calculated in \cite{chen} and
the best fitted stretched exponential function with $\beta =0.45$ and $\tau
=11.8$ $\mu $s.

\bigskip

Fig. 4 - Plots of the simulation result calculated in \cite{chen} (solid
line) and $I_{q}(t)$ (dotted line) with $q=1+(0.2)^{2}$ and $\alpha
=(3.61)^{2}\mu $s.

\bigskip

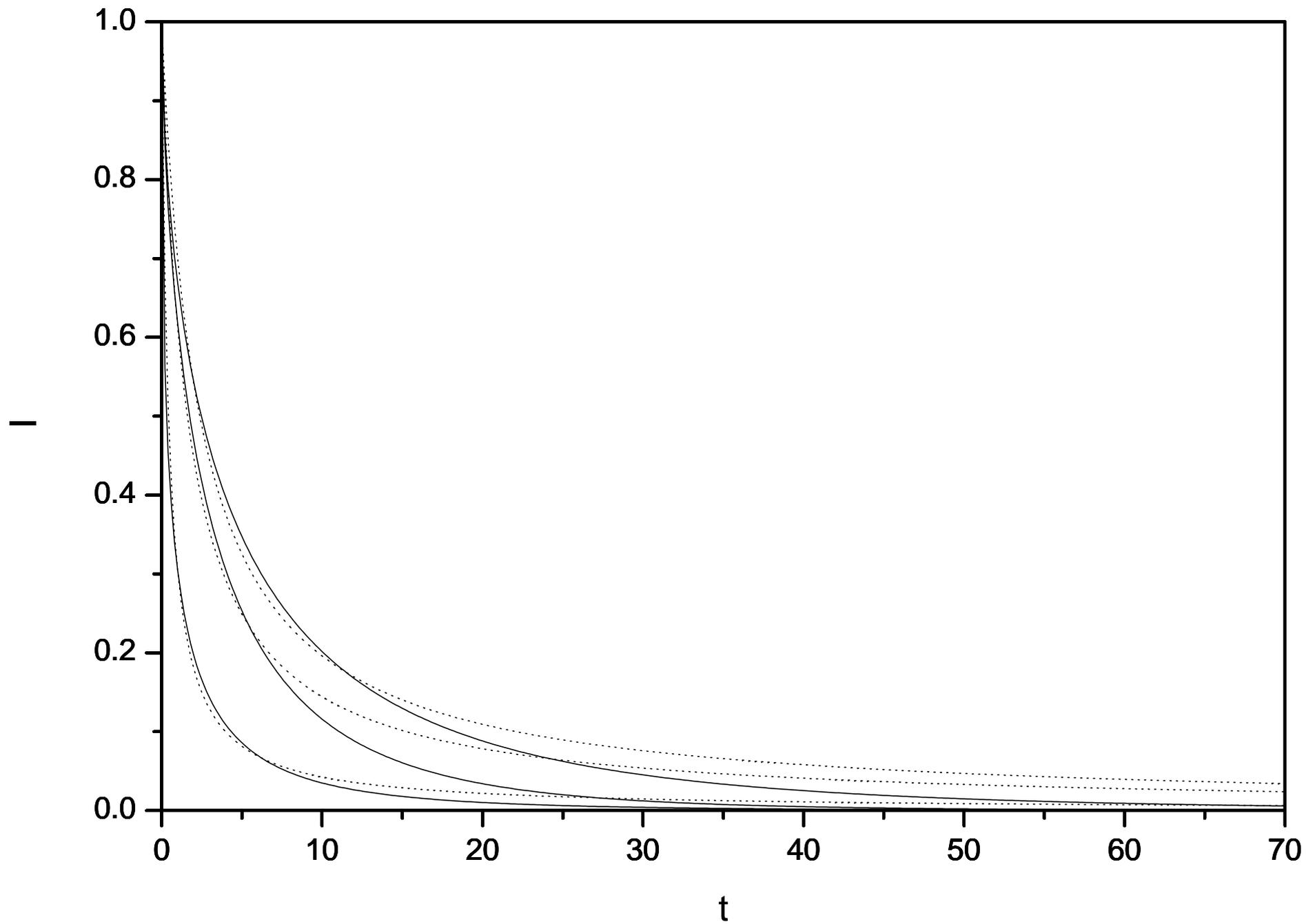
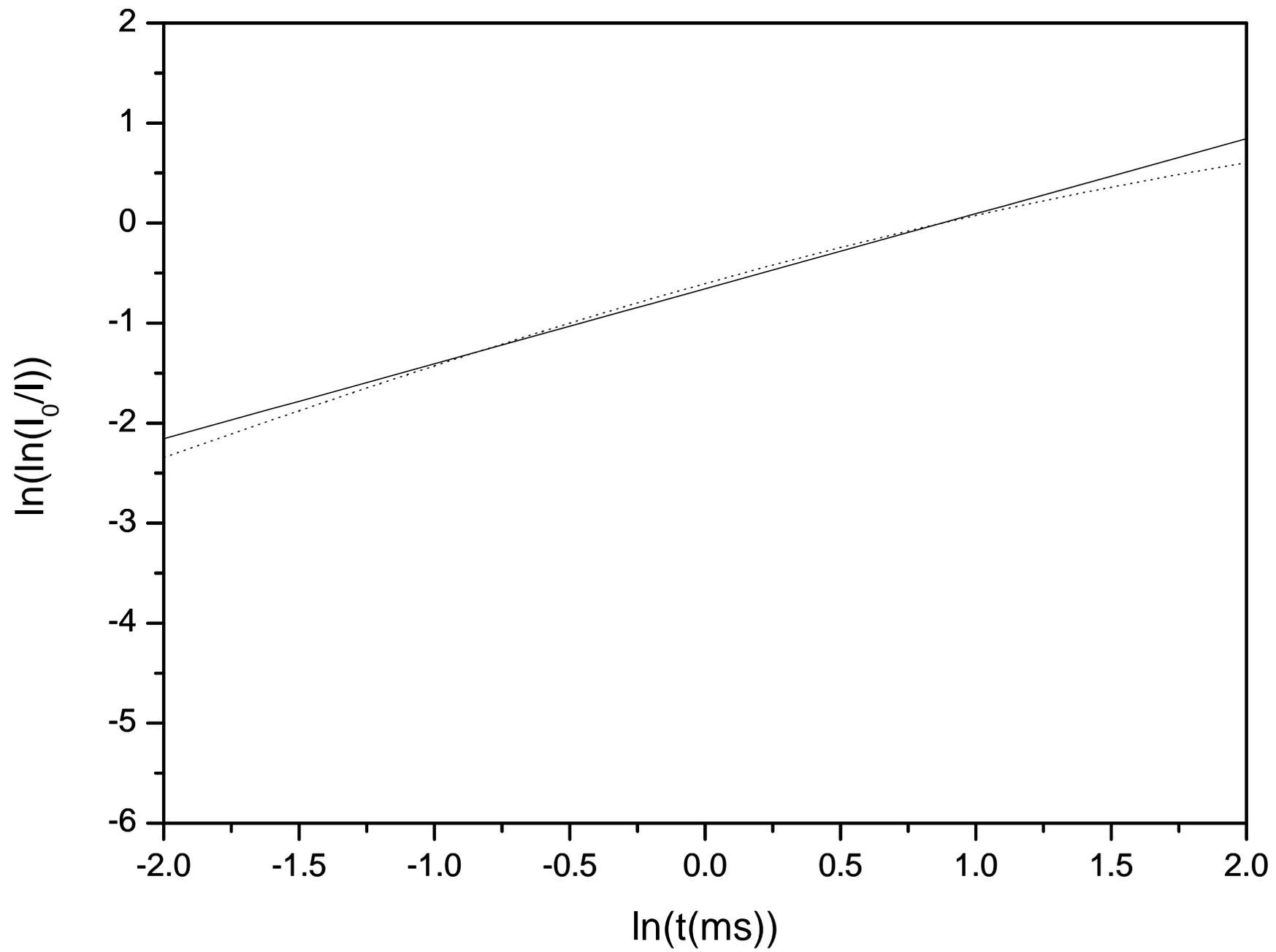
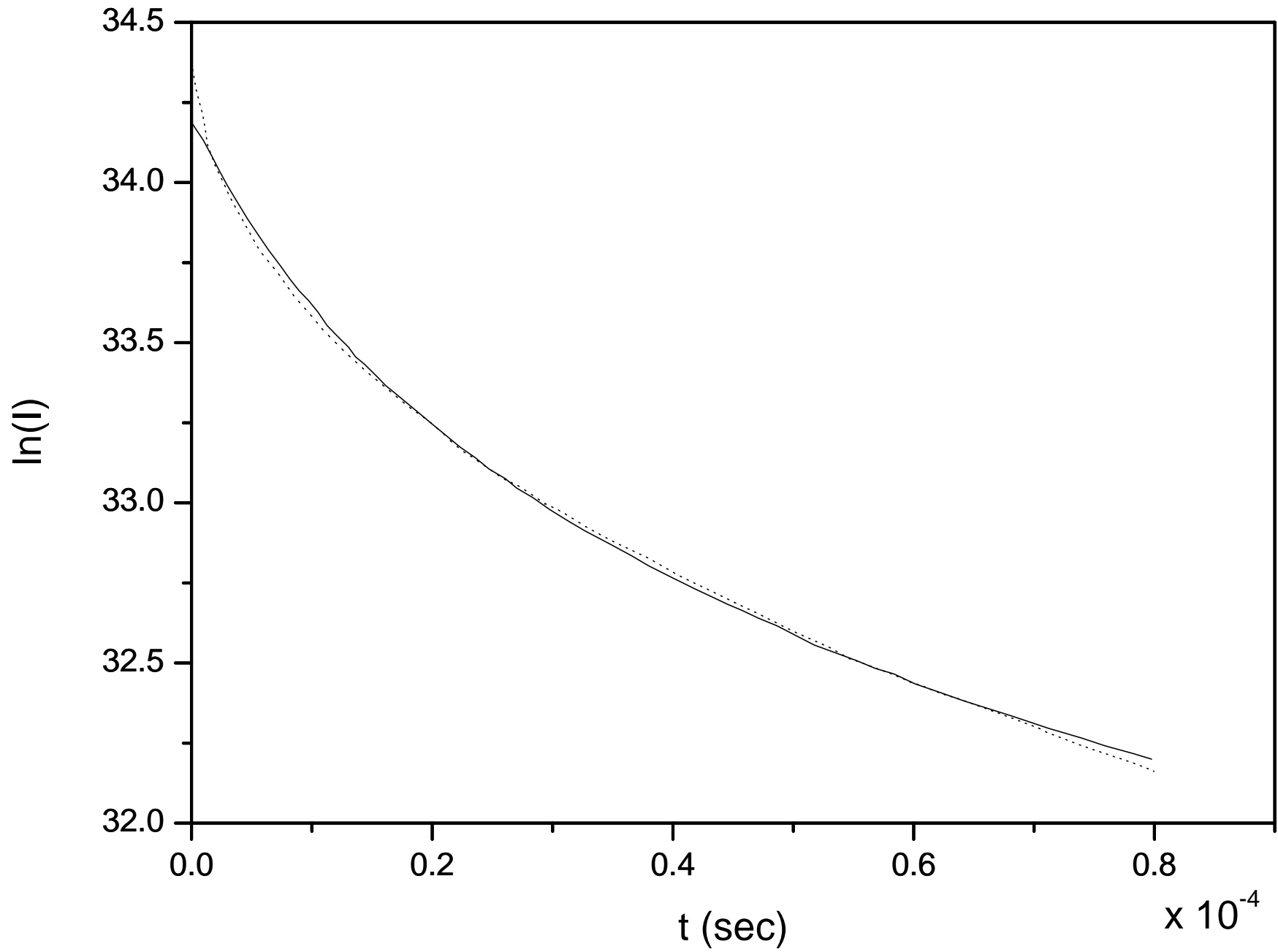
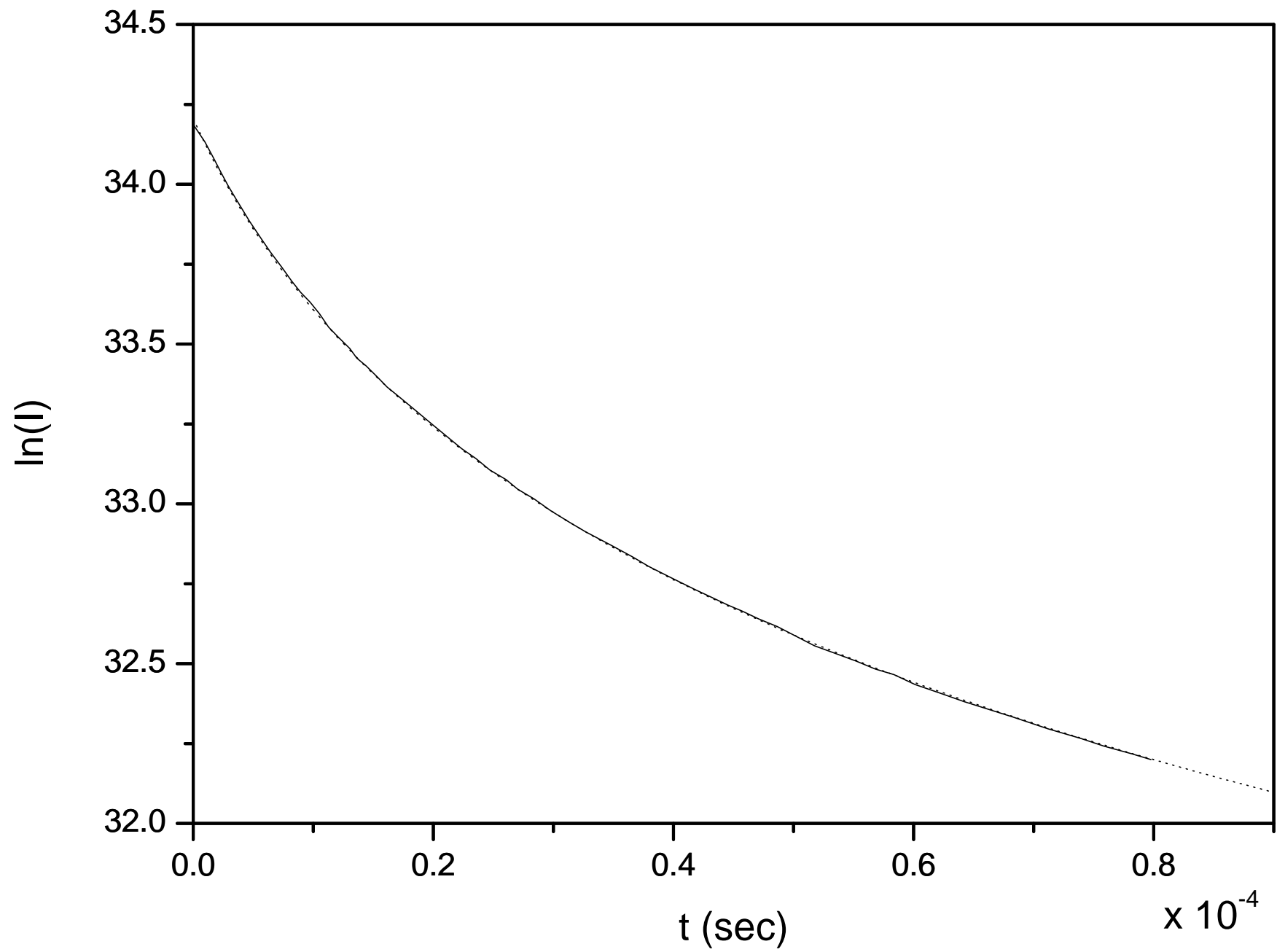
\end{document}